# Dynamic Cluster Head Node Election (DCHNE) Model over Wireless Sensor Networks (WSNs)

Abeer Alabass, Khaled Elleithy, and Abdul Razaque
Computer Science Department and Engineering
University of Bridgeport, CT, USA
(aalabbas, elleithy, arazaque)@bridgeport.edu

**Abstract**

WSNs are becoming an appealing research area due to their several application domains. The performance of WSNs depends on the topology of sensors and their ability to adapt to changes in the network. Sensor nodes are often resource constrained by their limited power, less communication distance capacity, and restricted sensing capability. Therefore, they need to cooperate with each other to accomplish a specific task. Thus, clustering enables sensor nodes to communicate through the cluster head node for continuous communication process.

In this paper, we introduce a dynamic cluster head election mechanism. Each node in the cluster calculates its residual energy value to determine its candidacy to become the Cluster Head Node (CHN). With this mechanism, each sensor node compares its residual energy level to other nodes in the same cluster. Depending on the residual energy level the sensor node acts as the next cluster head. Evaluation of the dynamic CHN election mechanism is conducted using network simulator-2 (ns2). The simulation results demonstrate that the proposed approach prolongs the network lifetime and balancing the energy consumption model among the nodes of the cluster.

## 1. Introduction

Wireless sensor networks experience limited communication bandwidth and energy constraints [1]. WSN is based on data-centric wireless network that does not require focusing on sender and receiver. Unlike, traditional wired network, mobile wireless network and ad-hoc network care more about the sender and receiver. Therefore, a general IP based mechanism and multi-hop routing scheme for mobile ad hoc network is not appropriate for WSNs [1]. The hierarchical routing protocol is one kind of typical network protocol for WSNs to handle the shortcomings of the flat traditional cluster-based routing scheme. Thus it extends the network lifetime as well as guarantees better connectivity of whole network [2].

Energy consumption is one of the serious problems in WSNs that creates challenges for academic and industrial sectors. Therefore, energy handling is one of the key skills to extend the network lifetime [3]. There is a quadratic increase in energy consumption as the distance among sensors increases [4]. Thus, the distance should be kept under consideration while designing the WSNs to minimize energy consumption and prolong network lifetime [5]. Scalability is the second major threat in WSNs [6] where thousands of sensor nodes are deployed in certain applications. In WSNs, these issues are addressed at the cluster level by using different cluster-based architectures. Leach algorithm introduced cluster-based architecture where the cluster head is static and its energy is consumed quickly due to the extra charge of data aggregation [7].

The cluster head selection and formation scheme has been introduced in [8] for handling the unstable energy depletion among self-powered sensor nodes within clusters. The cluster head in [8] decides to choose a timeframe on based on the energy consumed in the current round but it is probability based scheme. The dynamic clustering scheme for WSNs is introduced in [9] for focusing head node selection, cluster numbering and cluster reorganization for reducing energy consumption decided on signal strength and distance between the sink and the cluster head.

To control scalability, a clustering algorithm is proposed for selection of the cluster head node [10] based on the received signal strength of node and the distance between the cluster head node and the sink node. Another dynamic clustering algorithm is introduced in [11] using genetic algorithm. The algorithm considers different parameters to improve the network lifetime but most focusing on distance while selecting cluster head node. A cluster header selection scheme is proposed with new cluster formation mechanism in [12]. The node is selected as cluster head node on the basis of its relative energy consumption in the current round. This scheme tries to increase network lifetime and offers a balanced energy consumption configuration among the nodes.

The central node is selected as cluster head node by grouping the sensor nodes in cluster [13]. The purpose of this scheme is to reduce the energy consumption and improve the network lifetime. In this scheme is the cluster head dies quickly due to the static nature of the mechanism and it does not guarantee how much energy is available in the central node. To guarantee the stability in WSN, dynamic clustering Stable Sensor Network (SSN) is introduced to improve the efficiency of SSN based on heuristics and formulation of mathematical model [14]. The reported techniques in literature select the cluster head node either based on distance or relative energy of nodes.

In this paper we introduce the dynamic cluster head node election model that determines the energy of each node for selecting the cluster head node based on the maximum energy (residual energy) of sensor nodes in each cluster. This scheme reduces energy consumption and improves the lifetime of WSN. The reminder of the paper is organized as follows. In section 2, we describe our dynamic cluster head node election process. In section 3, the simulation and analysis of result are explained and section 4 provides a conclusion of the paper.

## 2. Dynamic Cluster Head Node Election process

Each sensor node has different energy level in its cluster at any given time. The energy level of the node depends on some factors such as sleep/wake up schedule, and amount of data transmitted and received. The sensor nodes are actively involved in detecting events and transmitting the information regarding the events. These activities lead to the death of the node faster than other nodes which are not actively involved. Under this condition, CHN experiences high energy consumption. To overcome this problem, our proposed mathematical model helps to determine the energy of each node in order to select CHN on the basis of maximum energy of sensor node in each cluster. Each sensor node determines its residual energy based upon consumed energy so far used in detecting events and transmitting its information.

This residual energy value determines whether the node should be considered as CHN candidate or not. The algorithm depends on calculating the residual energy of the sensor node and its distance from the base station if it is selected as CHN. The Non cluster head node (NCHN) detects CHN in its neighbor on the basis of multiple operations of WSNs using multiple rounds. The advantage of this approach is to provide enough flexibility to each NCHN to choose nearest CHN to reduce the energy consumption. Additionally, the process of choosing the CHN encompasses several steps. The base station broadcasts a short preamble message to each node of the WSN. Each node computes its distance from the base station based on the signal strength. The node that gets a short preamble message becomes a candidate CHN. each node waits until it gets an alert from another node of the cluster to compare its radio range and residual energy. If no message is received by another node that is supposed to be candidate node, then this node is elected as CHN. The elected CHN sends a multicasting message to its neighbor nodes in order to let them know about its election as the new CHN.

We determine the residual energy of each node in WSN on basis of a mathematical model. Let us assume that there is single-hop communication is used among sensor nodes to detect events and to transmit the information. Each node forwards data $'d'$ at distance 'r' within Cluster 'C' and located at A*A area of WSN. We determine the residual energy of two kinds of nodes: NCHN to CHN and CHN to Base station that can be expressed as follows.

$$R_{energy}(d,r) = \{d * E_{radio} + d * E_{amp}\frac{A^2}{2\pi C}\} \quad (1)$$

Where, $R_{energy}$: Residual energy of each node; $E_{radio}$: Energy consumption of radio; $E_{amp}$: Energy used for amplifying radio signal.

Equation (1) shows the residual energy of each node that sends data to CHN.

$$R_{energy}(d,r) = \{d * E_{radio} + d * E_{mh}r^4\} \quad (2)$$

Where, $E_{mh}$: Multi-hop fading channel.

Equation (2) shows the residual energy level of CHN when forwarding data to base station. Additionally, we need to determine the consumed energy when CHN communicates with NCHNs for receiving the data packets. It can be expressed as:

$$E_{RX(d)} = d * E_{radio}\left(\frac{S}{C} - 1\right) \quad (3)$$

Where, $E_{RX(d)}$: Energy consumed for receiving data packets and S: Number of sensor nodes.

CHN also consumes the energy in scheduling the cluster head sequence.

$$E_{RX(d)} = d * E_{schd}\left(\frac{S}{C} - 1\right) \quad (4)$$

$'E_{schd}'$ represents the energy consumed for scheduling. We hereby define three types of short preamble messages: $'d_{adv}'$, $'d_{syn}'$ and $'d_{join}'$ that show the short preamble message for advertisement, synchronization and joining the cluster respectively. Thus, CHN and NCHN consume energy during initial set-up phase that can be obtained as follows.

$$E_{CHN} = E_{T_X}(d_{adv}, r) + E_{R_X}(d_{adv}) + E_{T_X}(d_{syn}, r) \\ + E_{R_X}(d_{syn}) + E_{T_X}(d_{join}, r) \\ + E_{R_X}(d_{join}) \quad (5)$$

From equation (5), we deduce the consumed energy by CHN during setting initial phase of synchronization.

$$E_{NCN} = E_{T_X}(d_{adv}, r) + E_{R_X}(d_{adv}) + E_{T_X}(d_{syn}, r) \\ + E_{R_X}(d_{syn} + d) * E_{radio} \quad (6)$$

After the initial phase setting, NCHN and CHN nodes start to send data. On completion of event task, the final residual energy decides the election of next CHN can be expressed by equation (7) and (8).

$$E_{d_f\_CHN} = E_{d_f\_CHN} + [\, E_{R_X}(d_{size} + 1) - E_{R_X}(d_{size}) \\ + E_{R_X}(d_{size} + 1) \\ + E_{T_X}(d_{size} + 1) \_ E_{T_X}(d_{size}) \\ + E_{T_X}(d_{size} + 1) \quad (7)$$

The final residual energy of CHN can be determined using equation (7) that will decide CHN should remain as CHN or a new one is selected.

$$E_{d_f\_NCN} = E_{T_X}(d, r) \\ + [\, (d_{size} * E_{radio} + E_{T_X}(d_{size} + 1, r) \\ + E_{T_X}(d_{size} + n, r) \\ + E_{R_X}(d_{size} * E_{radio} + E_{R_X}(d_{size} \\ + 1, r) + E_{R_X}(d_{size} + n, r)] \quad (8)$$

Where, $d_{size}$ represents the size of data to be transmitted in each data packet and $'n'$ shows the number of packets transmitted and received by NCHN. Equation (7) and (8) decide the new election process of cluster head node.

## 3. Simulation and Analysis of Results

The evaluation of the proposed DCHNE scheme is conducted by using ns2 with Red Hat Enterprise Linux 6.5. We have compared DCHNE with three other known schemes, self-incentive and semi re-clustering (SISR) [12], low-energy, adaptive, clustering, hierarchy (LEACH) scheme [15], and round-robin clustering hierarchy (RRCH) scheme [16]. We have used 1 to 200 sensor nodes but the participating sensor nodes are 190. Sensor nodes are randomly placed at on area of 350m x 350m. The average distance between CHN and BS is set to 100m. We assume that each sensor node has 3.5 Joule initial energy and each node knows the location of other nodes using the algorithm of [15]. The simulation parameters are given in Table 1.

We have used two scenarios in this experiment. In the first scenario, obstacles are always available on remote area where sensor nodes are randomly deployed. Thus NCHN always send data to their CHN. In Scenario 2, obstacles are randomly available on remote area and the events are randomly generated over time. Thus, each node possesses different frequency for sleep/wake modes and different amount of transmitted/received data. The proposed DCHNE enables the nodes to utilize energy resources efficiently. In our experimental simulation, we have analyzed the network lifetime.

Figure 1 shows the network lifetime by displaying the number of alive nodes versus the set of time steps (frames) for four different schemes. The DCHNE delivers 634 to 1367 more packets than SISR, RRCH and LEACH. DCHNE outperforms the other three schemes due to the minimum overhead on each sensor node. After frame number 10648, we observe that five nodes are still alive which can take responsibility to become CHN. There is no need to send a message to those sensor nodes to determine their residual energy level, as this process also saves energy and prolongs network lifetime.

In Figure 2, we illustrate the results for a number of alive nodes as a function of time steps (frames). In scenario 2, DCHNE allows the nodes to forward the data packets only when they need to save energy for long distance transmission from CHN to BS. DCHNE also outperforms to other competing schemes. Furthermore, DCHNE delivers from 1623 to 31034 more data packets because the nodes are alive for longer time as compared with other schemes.

Table 1. Simulation Parameters

| Parameters | Description |
|---|---|
| Network size | 350m x 350m |
| Number of sensor nodes | 1 to 200 |
| Participating sensor nodes | 190 |
| Initial energy of sensor node | 3.5 joule |
| Data packet size | 500 bytes |
| Energy receive( $ER_X$ ) | 40 nJ/bit |
| Energy transmit( $ET_X$ ) | 40 nJ/bit |
| Simulation time | 20 minutes |
| Energy dissipation for free space | 9 pJ/bit/m$^2$ |
| Energy dissipation for two ray model | 0.0011 pJ/bit/m$^2$ |
| Energy aggregation data | 6 Nj/bit/message |
| Medium access control | BN-MAC [17] |
| Routing protocol | Energy aware routing protocol [18, 19] |
| Mobility | 2 m/sec |

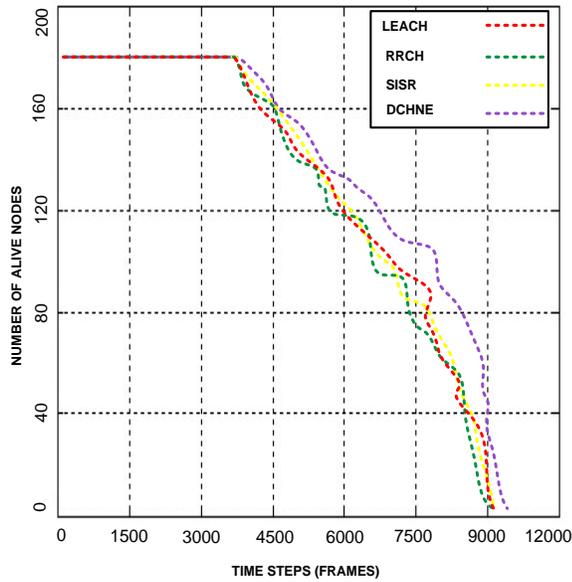

Figure 1: The number of alive sensor nodes VS time steps (frame) in scenario-1

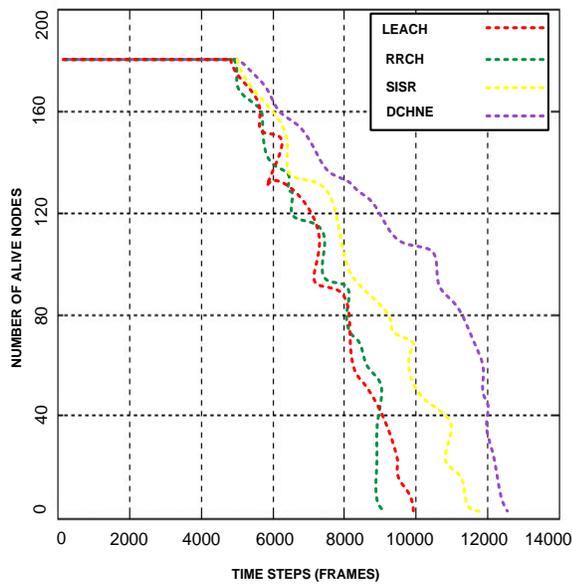

Figure 2: The number of alive sensor nodes VS time steps (frame) in scenario-2

## 4. Conclusions

In this paper, we have introduced dynamic cluster head node election (DCHNE) model over WSNs to prolong the network lifetime. We have shown dynamic election process of cluster head node. The cluster head node is elected on the basis of residual energy of sensor nodes. The residual energy is calculated after performing the event monitoring process using the mathematical model. In our scheme, the nodes can switch to their choice of cluster even with increased power loads.

To demonstrate the strength of DCHNE, we have used ns2 simulator that illustrates its performance efficiency. Two types of scenarios are used by algorithm which is characterized by the amount of activity perceived in the environments. On basis of simulation results and the mathematical model, we believe that the proposed scheme significantly prolongs the network lifetime as compared with other schemes.